\documentclass[12pt]{article}
\usepackage{amsmath}
\everymath{\displaystyle}
\newtheorem{theorem}{Theorem}

\newtheorem{corollary}{Corollary}

\newtheorem{lemma}[theorem]{Lemma}
\newtheorem{proposition}{Proposition}

\begin{document}

\title{Stability Number and f-Factors in Graphs}

\author{Mekkia KOUIDER}

\maketitle

km@lri.fr

\begin{abstract}
Let $f: X \longrightarrow  N $ be an integer function.
An $f$-factor is a  spanning subgraph of a graph $G=(X,E)$ whose vertices have 
degrees defined by $f$ .
 In this paper, we prove a sufficient condition for
the existence of a $f$-factor which involves the stability number, the minimun
degree of $G$ or the connectivity of the graph.
\end{abstract}

\vspace{4mm}

\vspace{4mm}

\section{Introduction}\label{intro}

We consider simple graphs without loops. For notation and graph theory 
terminology we follow in general \cite{w}.
Let $G$ be a graph with vertex set $X$ and edge set $E(G)$. Denote by 
$d_{G}(x)$ the degree of a vertex $x$ in $G$,
and by $\delta(G)$ the minimum degree of $G$. A \emph{spanning subgraph} of 
$G$ is a subgraph of $G$ with vertex set $X$.
Let $f: X \longrightarrow  N $ be an integer function.
For any subset $A$ of $X$, we denote by $f(A)$ the sum $\sum_{x \in A} f(x).$
A spanning subgraph $H$ of a graph $G$ such for every vertex $x$, 
$d_H (x)= f(x),$ is called an $f$-factor of $G$.
 Let $a, b$ be fixed integers. A spanning subgraph $F$ of $G$ is called an 
$[a,b]$-\textit{factor} of $G$ if $a\leq d_{F}(x)\leq b$ for all $x\in X$.

For $S\subseteq X$, let $|S|$ be the number of vertices in $S$ and let $G[S]$ be 
the subgraph of $G$ induced by $S$.
We write $G-S$ for $G[X\backslash S]$. A  set $S\subseteq X$ is 
called independent if $G[S]$ has no edges.
Denote by $\alpha(G)$ the stability number of a graph $G$, by $\kappa(G)$ 
its vertex connectivity.
 For any vertex $v \in X$,
the \textit{open neighborhood} of $v$ is the set 
$N(v)=\{u \in X\backslash uv\in E(G)\}$; for a set $A\subseteq X$,
$N_{G}(A)$ denotes the set of neighbors in $G$ of vertices in $A$. 
Given disjoint subsets $A, B \subseteq X$, we write
$e(A,B)$ for the number of edges in $G$ with one extremity in $A$ and 
the other one in $B$.\\
If $S$ is a cutset, let $h'(G-S)$ be the number of components $C$ of $G-S$ such 
that $\sum_{x \in C} f(x)$ is odd.

Let $t$ be a nonnegative real number.
 We say that $G$ is $t$ odd-tough if for each cutset $S$,
$h'(G-S) \leq |S|/t$.
We remark that if $G$ is $t$ tough then $G$ is $t$ odd-tough.
\section{Known Results}

Given a graph $G=(X,E)$, an application $f$ and a cuple of
disjoint subsets of $X$, we recall that an {\it odd component} C of 
$G- (S \cup T)$ is a component $C$ such that 
odd.\

Many authors have investigated $f$-factors , see for example  \cite{v}.
Tutte (\cite{l}) gave the well-known necessary and sufficient condition 
for existence of an $f$-factor.\

\noindent
\bf{Condition} \cite{t} 
\it{A graph $G=(X,E)$ has an $f$-factor if and only if\

1)
$\delta(S,T)= f(S)-f(T)+\sum_{v \in T} d_{G \backslash {S}}(v) -h(S,T)\geq 0$, 
disjoint subsets $S$ and $T$ of $X$\

where $h(S,T)$ is the number of odd components of $G-(S\cup T)$\

2) $\delta(S,T) \equiv f(X) (mod 2)$.}

\vspace{5mm}

\rm
This condition is also a corollary of the $(g,f)$ factor theorem of Lov\'{a}sz in \cite{l}. 
However, in practise, this condition remains difficult to verify.

\vspace{5mm}

Katerinis and Tsikopoulos established a condition on the minimum degree for 
the existence of $f$-factors.

\begin{theorem} \emph{\cite{kat2}}
 Let $b\geq a$ two positive integers and let $G=(X,E)$ be a graph 
with the minimum degree $\delta$. \
Suppose $\delta \geq \frac{b.|X|}{a+b}$, and $|X| > {(a+b)(b+a-3)}/a.$
If $f$ is a function from $X$ to $\{a,a+1,...,b \}$ such that $f(X)$ is even,
then $G$ has an $f$-factor.
\end{theorem}

In \cite{kat1}, Katerinis has a condition on the toughness of the graph.

\vspace{5mm}

 Only few results are known which relate the stability number and factors.
Nishimura had sufficient condition for a k factor.

\begin{theorem} 
Let $r \geq 1$ be an odd integer, and $G$ be a graph of even order. of 
connectivity $\kappa$.
If $\kappa \geq {(r+1)^2} /2$,and, 
$\alpha(G) \leq \frac{4r. \kappa}{(r+1)^2}$,
then  $G$  has  an  $r$-factor.
\end{theorem}

\vspace{2mm}
\rm
The following result involving the stability number and the minimum
degree of a graph was given by M. Kouider and Zbigniew Lonc \cite{lo}:\\
\vspace{2mm}

\begin{theorem} \emph{\cite{lo}}
 Let $b\geq a+1$ and let $G$ be a graph with the minimum degree
$\delta$. 
If $\alpha(G) \leq 4b(\delta-a+1))/(a+1)^2$, for $a$ odd\
   and $\alpha(G) \leq 4b(\delta-a+1)/a(a+2)$, for $a$ even.\
                             
then  $G$  has  an  $[a,b]$-factor.

\end{theorem}

Cai has shown that 
\begin{theorem} \emph{c}
Let $G$ be a connected $K_{1,n}$-free graph and let $f$ be a nonnegative 
integer-valued fonction on $V(G)$ such that 
$1 \leq n-1 \leq a \leq f(x) \leq b$ for every $x \in V(G)$.\

If $f(V(G))$ is even , $\delta(G)\geq b+n-1$
and 
$\alpha(G) \leq \frac {4a.(\delta- b-n+1)}{(n-1)(b+1)^2}$,
then $G$ has an $f$ factor.
\end{theorem}

Note that Cai conjectured that that the condition on the stability
$\alpha(G) \leq \frac {4a.(\delta- b)}{(b+1)^2}$ is sufficient in connected graphs.
We have the following counterexample.
\vspace{3mm}
\rm

Suppose $b$ is an odd integer and $a$ an integer strictly less than $b$.

Let $G_0$ be a connected graph of minimum degree $\delta$ at least $(b+1)^3 +b$.
Let    $p= \frac {4a.(\delta- b)}{(b+1)^2}$. 
In the graph $G_0$, we suppose there exists  $S$ be a  cutset of $k< b$ 
vertices, such that $G(S)$ is complete and $C_1,...,C_p$, the connected components of $G-S$,
 form a family of complete subgraphs of order $\delta +1$,  
mutually independent. Furthermore $G(S \cup C_1)$ is complete, and,
for each $ i \geq 2$, exactly one edge joins $S$ to  $C_i$. 
So  $\alpha(G_0)=p= \frac {4a.(\delta- b)}{(b+1)^2}$.\
 
\vspace{3mm}
Let us consider the application $f$ on $X$ such that \
$f(x)= a$ if $x \in S$, and, $f(x)= b$ otherwise.\

If a $f$ factor exists we should have 
$\alpha=c(G-S) \leq a.k$, so $c(G-S)$ should be at most $ab$.
This is not satified  as 
$\alpha= \frac {4a.(\delta- b)}{(b+1)^2} > 4a(b+1)$.

\vspace{5mm}
One can see the surveys \cite{p} or  \cite{v} for other results.

\vspace{5mm}
\section{Main Results}

We have established a new sufficient condition for
a graph  to have an $f$-factor; this condition involves the stability number, 
the minimum degree of the graph.\

\begin{theorem}
Let  $b \geq 2$  be an integer and let $G=(X,E)$ be a connected graph, 
of minimum degree $\delta$ at least $b$.
Let $f$ be a non-negative integer valued function on $X$, such
 that for each $x \in X$, $a \leq f(x) \leq b$ and $f(X)$ is even.
If $\alpha(G) \leq \frac {4a.(\delta- b)}{(b+1)^2}$, 
and the odd-toughness of $G$ is at least $1/a$, then \

 $G$ contains an  $f$-factor.
\end{theorem}

\vspace{5mm}

Furthermore, we get this corollary.
 \begin{corollary}
Let  $b \geq 2$  be an integer and let $G=(X,E)$ be a graph, of 
minimum degree $\delta$ at least $b$ and connectivity $\kappa$.
Let $f$ be a non-negative integer valued function on $X$, such
 that for each $x \in X$, $a \leq f(x) \leq b$ and $f(x)$ is even.
If $\alpha(G) \leq  \frac {4a.(\delta- b)}{(b+1)^2} $, 
then \

 $G$ contains an  $f$-factor.
\end{corollary}

\vspace{4mm}

\begin{corollary}
Let  $b\geq 2$  be an integer and let $G=(X,E)$ be a graph, of 
minimum degree $\delta$ at least $b$ and connectivity $\kappa$.
Let $f$ be a non-negative integer valued function on $X$, such
 that for each $x \in X$, $a \leq f(x) \leq b$ and $f(X)$ is even.
If $\alpha(G) \leq  min( \frac {4a.(\delta- b)}{(b+1)^2}, a \kappa )$, 
then \

 $G$ contains an  $f$-factor $\bullet$
\end{corollary}

\rm
\vspace{4mm}

The condition $\alpha(G) < \frac{4a.(\delta- b)}{(b+1)^2}~~+~1 $ is 
necessary if $b >2a$.
Let  $\alpha > \delta > b > r$ be four integers.
Let us consider a graph $G_1$ composed by the join of a  complete graph 
$A=K_{\delta-r+1}$ and $B$, the disjoint union of $\alpha$ complete graphs of
order $r$.
Let $f$ be a function such that 
\vspace{2mm}

$f(x)=a$ if $x \in X(A)$, $f(x)=b$ if $x \in X(B)$.
If an $f$ factor exists we get 
$$\alpha(G) \leq \frac{a.(\delta- r+1)}{r.(b+1 -r)}~.$$
For $b$ odd and $r= (b+1)/2$, we get 
$\alpha(G) < \frac{4a.(\delta- b)}{(b+1)^2}~+~\frac{2a}{b+1}.$

\vspace{4mm}

\section{Proof of Theorem 4}
We set first some usefull lemmas.

\begin{lemma}
 $\delta(S,T)$ is even.
\end{lemma}

{\it Proof}
Let ${\mathcal I}_1$ (respectively ${\mathcal I}_2$) be the set of even (resp.
odd) components of $G- (S\cup T)$.
By definition,
$$f({\mathcal I}_1)\equiv e({\mathcal I}_1,T),~~ (1)$$
$$f({\mathcal I}_2)\equiv h(S,T) +  e({\mathcal I}_2,,T),~~~(2)$$
so, by (1) and (2),
 $$f(X)= f(S)+f(T)+ f({\mathcal I}_1 + f({\mathcal I}_2)
\equiv f(S) -f(T) + e(G-(S\cup T),T) + h(S,T).$$
As $f(X)$ is even, the conclusion follows.

\vspace{5mm}

\begin{lemma}
 $T$ is non-emptyset.
\end{lemma}

{\it Proof}

 If $T=\emptyset$ and $S= \emptyset$, then $\delta(S,T)=-h=0$ as 
$G$ is connected and $f(X)$ is even.
If $T=\emptyset$ and $S$ is not empty,
then $h(S,T)$ is the number of components of $G-S$ such that $f(C)$ is odd.

Either $S$ is not a cutset, then $h(S,T) \leq 1\leq a|S|$;
or $S$ is a cutset, as $G$ is $1/a$-tough, $h \leq a|S|$.

As $a|S| \leq f(S) $, 
then $\delta(S,T)=f(S) -h(S,T) \geq f(S)-a|S| \geq 0$.


\vspace{5mm}

\begin{proposition}
If $\alpha(G) \leq \frac {4a.(\delta- b)}{(b+1)^2}$, then \
$$|S| > \delta-b$$

\end{proposition}

{\it Proof}

The proof is by contradiction.
As $\delta(S,T) < 0$ and $a \leq  f(x) \leq b$ for each $x$, 
then
$$ ( \delta - |S|)|T| + a|S|-b|T|-h<0,$$
so $$ ( \delta - |S| -b)|T| <h- a|S|.$$
If $|S| = \delta-b$, we get $|S|< \frac{h}{a} < \frac{\alpha(G)}{a}<\frac {4 (\delta-b)}{9}.$
This a contradiction.\

Now we assume $|S| < \delta-b$, and we get
$$ |T| < \frac{h- a|S|}{( \delta - |S| -b)}.$$

If $h<a |S|$, then $|T|=0$.
As $h< \alpha$, then
$$|T| < \frac{4a}{(b+1)^2}. 
\frac{(( \delta - |b|)-{(b+1)^2}|S|  )}{( \delta - |b|-|S|)}$$
We get
$$|T| < \frac{4a}{(b+1)^2}. 
( 1 - \frac{{({(b+1)^2}/4} ~-1).|S|}{( \delta - |b|-|S|)})$$
$$T < \frac{4a}{(b+1)^2}.$$
As $b \geq a $,
$T < \frac{4a}{(b+1)^2} \leq 1$, so $|T|=0$. This is in contradiction 
with Lemma 5.

\vspace{1cm}
{\it End of the proof of the theorem}

\vspace{5mm}
Let $h_2$ be the number of components of $G -(S\cup T)$ not 
adjacent to $T$.
As $\delta(S,T)<0$, we have

$$2|E_T|+|T|+a|S| -(b+1)|T| -h_2 \leq 0,~~~(1) $$ 

As $\alpha_T$ the stability number of $T$ is at least 
$\frac{|T|^2}{2|E_T|+|T|}$,  and \

$\alpha_T \leq \alpha (G)-h_2$,
we get, using (1), \

$$\alpha (G)-h_2 \geq \frac{|T|^2}{(b+1)|T|-a|S|+h_2}$$

Let us set $|T|= r.|S|$.
Then 
$$\alpha (G)-h_2 \geq \frac{r^2.|S|^2}{(b+1)r|S|-a|S|+h_2}$$
$$\alpha (G)-h_2 \geq \frac{r^2.|S|}{(b+1)r -a +{h_2}/|S| }$$
The minimum of the bound as a function of $r$ is for 
$r=\frac{2(a- {h_2}/|S|)}{b+1}.$ \
It follows that 
$$\alpha (G)-h_2 \geq \frac{4a.|S|}{(b+1)^2}-\frac{4h_2}{(b+1)^2}$$
As by hypothesis $\alpha(G) \leq \frac{4a (\delta-b)}{(b+1)^2}$,
and $|S| \geq  (\delta-b)$, we get \

$$h_2 \leq \frac{4h_2}{(b+1)^2}.$$
So $h_2=0$. As $\delta \geq b$, then $\delta(S,T) \geq 0$.
This is a contradiction with the definition of the cuple $S,T$.

This ends the proof of the theorem 4 $\bullet$

\end{document}